\documentclass{article}
\usepackage{spconf}
\usepackage[utf8]{inputenc}
\usepackage[T1]{fontenc}
\usepackage{amsmath,amsfonts}
\usepackage{graphicx}
\usepackage{hyperref}
\usepackage{url}
\usepackage{booktabs}
\usepackage{svg}
\usepackage{nicefrac}
\usepackage{microtype}
\usepackage[table]{xcolor}

\newcommand{\vx}{\mathbf{x}}
\newcommand{\vy}{\mathbf{y}}

\newcommand{\balpha}{\boldsymbol{\alpha}}
\newcommand{\bbeta}{\boldsymbol{\beta}}
\newcommand{\bgamma}{\boldsymbol{\gamma}}
\newcommand{\score}[2]{%
\begin{tabular}{@{}c@{}}
$#1$\\[-1pt]
{\scriptsize\textcolor{gray}{$\pm #2$}}
\end{tabular}%
}

\newcommand{\bestscore}[2]{%
\begin{tabular}{@{}c@{}}
\textcolor{green!50!black}{$\mathbf{#1}$}\\[-1pt]
{\scriptsize\textcolor{gray}{$\pm #2$}}
\end{tabular}%
}

\newcommand{\secondscore}[2]{%
\begin{tabular}{@{}c@{}}
{$\underline{#1}$}\\[-1pt]
{\scriptsize\textcolor{gray}{$\pm #2$}}
\end{tabular}%
}

\title{SCLAP: Temporal Structural Alignment Improves Localization \\ in Long-Form Audio}
\name{Yacouba Kaloga, Ina Kodrasi\thanks{This work was supported by the Swiss National Science Foundation project CRSII5 202228 on “Characterisation of motor speech
disorders and processes”.}}
\address{Signal Processing for Communication, Idiap Research Institute, Martigny, Switzerland}
\begin{document}
\ninept
\maketitle
\begin{abstract}
Audio--text models such as Contrastive Language-Audio Pretraining (CLAP) typically pool frame and token sequences uniformly into global embeddings. This discards temporal structure and impairs both retrieval and localization, especially for short events in long recordings. We introduce SCLAP, a sequence-level contrastive framework that aligns audio frames and text tokens using monotone one-dimensional optimal transport. Its learned marginals and transport plan provide fine-grained temporal supervision from clip-level captions alone. On short-form audio (i.e., Clotho and TACOS datasets with 15-30s recordings), SCLAP remains competitive for retrieval while improving zero-shot localization over CLAP and Text-aware Attention Pooling. On long-form audio (i.e., CASTELLA dataset with up to 300s recordings), SCLAP improves retrieval performance and its localization advantage grows as the event-to-context ratio decreases. At inference, SCLAP requires no optimal transport computation since retrieval can use pooled embeddings and localization can use inexpensive frame--text cosine scores.

\end{abstract}
\begin{keywords}
language, audio, CLAP, retrieval, optimal transport
\end{keywords}

\section{Introduction}
\label{sec:intro}

Contrastive multimodal pretraining has become a standard way to learn shared
representation spaces across modalities, enabling zero-shot recognition and
transfer to downstream tasks~\cite{radford2021,jia2021}. Contrastive
Language-Audio Pretraining (CLAP)~\cite{elizalde2022claplearningaudioconcepts}
extends this recipe to audio--text pairs. Like other dual-encoder systems with
sequential or temporally structured inputs such as audio--text or
video--text learning, CLAP represents each modality as a sequence of embeddings
and then uniformly pools these representations into a \emph{single} global
vector~\cite{elizalde2022claplearningaudioconcepts,guzhov2021audioclipextendingclipimage}.
This recipe has proven remarkably effective, since it is simple to train, scales well
to large corpora, and achieves strong zero-shot classification and
cross-modal retrieval performance~\cite{wu2024largescalecontrastivelanguageaudiopretraining}.

Global pooling implicitly assumes that a single vector can adequately summarize the entire input. 
This assumption holds reasonably well on short-form audio benchmarks, where relevant events often occupy a substantial portion of the input and transformer-based encoders can aggregate local context through self-attention.
However, in long-form audio, recordings are often processed chunk by chunk while the captioned event may occupy only a small fraction of the timeline. 
In this low \emph{event-to-context ratio} regime, uniform global pooling can dilute the relevant event among background, silence, or irrelevant context.
The event's presence can often still be inferred from the pooled vector, however, its temporal location within the clip is lost~\cite{oncescu2024dis}. 

Such temporal localization is essential particularly in long recordings, where retrieving relevant segments and navigating directly to them requires knowing not just that an event occurred, but also when.
Existing approaches often recover temporal information using additional supervision. Examples include temporal-contrastive
objectives~\cite{yuan2024tc}, frame-wise training with temporally grounded
captions~\cite{primus2025tacos}, and supervised audio moment
retrieval~\cite{munakata2025lan}. Related image--text methods use auxiliary
objectives~\cite{zhao2025superclip}, while~\cite{li2026fine} relies on frame-level labels. Although effective, these
approaches require temporal annotations beyond ordinary clip-level captions.

In contrast, comparatively few methods learn fine-grained audio--text structure using only clip-level supervision. Li et al.~\cite{li2024mga} align frames and tokens through a shared codebook, but the approach is tested only on short audio clips.
For long-form audio, Contrastive Long-form Language-Audio Pretraining (CoLLAP)~\cite{wu2024collap} applies temporal attention over audio segments with respect to a global caption embedding for (long-form) music retrieval, but this mechanism is never evaluated for its ability to localize events in time. %Contrastive Long-form Language-Audio Pretraining (CoLLAP)~\cite{wu2024collap} computes temporal attention between audio frames and text tokens internally for music retrieval, but this mechanism is never evaluated for its ability to localize events in time.
A related direction is Text-aware Attention Pooling (TAP)~\cite{xin2023improvingtextaudioretrievaltextaware} which lets a single caption-level query attend over audio frames, allowing the model to emphasize the most relevant audio frames. However this produces a single attention distribution for the whole caption, without an alignment between individual text tokens and audio frames. The resulting weights are never supervised or evaluated as a localization signal, and are only used to build a better pooled vector for retrieval.

These limitations motivate the necessity of a mechanism that aligns frames and tokens directly without collapsing either sequence first.
Optimal transport~\cite{peyré2020computationaloptimaltransport} offers a natural way to obtain such a correspondence. Rather than collapsing two sequences into single vectors before comparing them, optimal transport finds a soft assignment between individual frames and individual tokens that minimizes the total cost of aligning them. 
Although prior work has applied optimal transport to audio-text retrieval, it has only been used to match pooled representations of different examples across a training batch~\cite{luong2024revisiting,pmlr-v235-shi24b} rather than to align the frames and tokens within a single pair.

In this paper we propose \textbf{S}equence-level \textbf{C}ontrastive \textbf{L}anguage--\textbf{A}udio \textbf{P}retraining (\textbf{SCLAP}), which applies optimal transport within each audio--caption pair to align its frame and token sequences directly. A lightweight network predicts an importance weight for each audio frame, and a separate lightweight network predicts an importance weight for each text token, using only that modality's own sequence. These weights induce a monotone transport plan between the frames and tokens, and SCLAP replaces CLAP's pooled similarity with the resulting transport-weighted similarity in the same contrastive objective. Because this objective rewards weights that help distinguish matched from mismatched pairs, training shapes them into meaningful importance scores without any supervision beyond the ordinary clip-level caption. Since each modality's weights are computed independently of the other, SCLAP does not need the transport plan at inference. Instead, for retrieval SCLAP can pool each modality into a single embedding using its own learned weights and compare the two pooled embeddings directly with cosine similarity. For localization, it instead compares each individual audio frame against the pooled text embedding, yielding a per-frame score that shows where in the audio the caption's content is concentrated.
Experimental results show that on short-form audio (i.e., on Clotho~\cite{drossos2019} and TACOS~\cite{primus2025tacos} datasets), SCLAP remains competitive with pooling- and attention-based models such as TAP for global retrieval, while improving zero-shot localization. On long-form audio (i.e., on the CASTELLA dataset~\cite{munakata2025castella}), SCLAP obtains the strongest mean retrieval performance and localizes events more accurately than global pooling and TAP, with its localization advantage increasing as more irrelevant context surrounds the target event. 
Our main contributions are:

\begin{itemize}
\item \textbf{Token-level contrastive alignment:}
We introduce SCLAP, a temporal optimal-transport objective that aligns
audio-frame and text-token sequences using only clip-level supervision.

\item \textbf{Improved long-form localization:}
SCLAP localizes described events more accurately than mean pooling and
TAP on long recordings, with its advantage increasing
as more irrelevant context surrounds the event.

\item \textbf{Transport-free inference:}
We empirically show that SCLAP retains its gains using marginal-weighted
pooling for retrieval and frame--text cosine scores for localization, without needing to solve optimal transport at inference.
\end{itemize}

\section{Method}
\label{sec:method}

We first review CLAP-style contrastive learning (Section~\ref{sec:clap-review}) and the Sequence Optimal Transport Distance (SOTD) framework underlying SCLAP (Section~\ref{sec:sotd}), before introducing our sequence-level contrastive similarity (Section~\ref{sec:sclap-similarity}). 
\subsection{CLAP-style contrastive learning}
\label{sec:clap-review}
Let $\{\vx\}_n=(\vx_1,\dots,\vx_n)$ denote a sequence of audio-frame embeddings and
$\{\vy\}_m=(\vy_1,\dots,\vy_m)$ denote a sequence of text-token embeddings. CLAP-style models pool
each sequence into a single $\ell_2$-normalized global embedding
$\bar{\vx}$ and $\bar{\vy}$ and compare them using the temperature-scaled cosine
similarity $s(\bar{\vx},\bar{\vy})=\bar{\vx}^\top\bar{\vy}/\tau$, where $\tau$ is the contrastive temperature.
Given a batch of $B$ audio-text pairs with similarity $s_{ij}=s(\bar{\vx}_i,\bar{\vy}_j)$, where $i=1, \ldots, B$ and $j = 1, \ldots, B$, CLAP minimizes
the symmetric InfoNCE loss~\cite{elizalde2022claplearningaudioconcepts}
\begin{equation}
\label{eq: infonce}
	\mathcal{L}_{\mathrm{CLAP}} = -\frac{1}{2B}\sum_{i=1}^{B}
	\left(
	\log\frac{e^{s_{ii}}}{\sum_j e^{s_{ij}}}
	+
	\log\frac{e^{s_{ii}}}{\sum_j e^{s_{ji}}}
	\right).
\end{equation}
Typically, $\bar{\vx}_i$ is the mean-pooled audio representation whereas
$\bar{\vy}_j$ is the text \texttt{[CLS]} token.

\subsection{Sequence Optimal Transport Distance}
\label{sec:sotd}

We have introduced the Sequence Optimal Transport Distance (SOTD) in~\cite{kaloga2025differentiable} to compare two ordered sequences  for automatic speech recognition (ASR) by transporting mass between their elements, while preserving the order of both. Let $\balpha\in\Delta^n$ and
$\bbeta\in\Delta^m$ be normalized weights (with a softmax) on the audio frames and text tokens. Viewing
each sequence as a discrete distribution on an ordered one-dimensional
support, the pair $(\balpha,\bbeta)$ admits a unique monotone coupling
$\bgamma(\balpha,\bbeta)$ obtained by matching their cumulative distributions in $\mathcal{O}(n+m)$. Given a pairwise cost $C$, the SOTD is defined as~\cite{kaloga2025differentiable}
\begin{equation}
\mathcal{S}_r(\vx,\vy)
=
\min_{\balpha\in\Delta^n}
\Big(
\sum_{i,j}\gamma_{ij}(\balpha,\bbeta)\,C(\vx_i,\vy_j)^r
\Big)^{1/r},
\label{eq:SOTD}
\end{equation}
with $r\in\mathbb{N}^{*}$. In the ASR setting, $\bbeta$ is fixed to the
uniform distribution, since every token of the transcript must be accounted
for, whereas $\balpha$ is optimized, allowing frames without content (e.g., silence) to receive no mass. 
To avoid solving the minimization at every training step, \cite{kaloga2025differentiable} predicts
$\balpha$ with a lightweight scalar-output network trained end-to-end.
SCLAP extends this
parameterization to both the audio and text marginals, similarly avoiding the SOTD optimization during training.

\subsection{SCLAP: Sequence-level Contrastive Language--Audio Pretraining}
\label{sec:sclap-similarity}
SCLAP adapts the SOTD framework of \cite{kaloga2025differentiable} to the
audio--text setting with three changes. First, we set $r=1$ in~(\ref{eq:SOTD}) and use the cosine cost $C(\vx_i,\vy_j)=1-\vx_i^\top\vy_j$.
Since $\bgamma(\balpha,\bbeta)$ sums to one, this lets us rewrite the SOTD
cost as $1-\sum_{i,j}\gamma_{ij}(\balpha,\bbeta)\,\vx_i^\top\vy_j$, so that
minimizing it is equivalent to maximizing the transport-weighted cosine
similarity $\sum_{i,j}\gamma_{ij}\vx_i^\top\vy_j$. Second, rather than fixing $\bbeta$ to the uniform
distribution, we predict \emph{both} marginals with separate lightweight
scalar-output networks $\mathbf{W}_x,\mathbf{W}_y$ and temperatures $\theta_x,\theta_y$, i.e.,
{\small
\begin{equation}
\balpha=\mathrm{softmax}\!\left(
\frac{[\mathbf{W}_x(\vx_i)]_{i=1}^{n}}{\theta_x}
\right),
\bbeta=\mathrm{softmax}\!\left(
\frac{[\mathbf{W}_y(\vy_j)]_{j=1}^{m}}{\theta_y}
\right).
\end{equation}
}
Third, since we never solve for $\balpha$ via the SOTD minimization, we use
the resulting quantity directly as a similarity score rather than as an
approximation to a metric distance, i.e.,
\begin{equation}
s_{\mathrm{SCLAP\text{-}T}}(\{\vx\}_n,\{\vy\}_m)
=
\sum_{i,j}\gamma_{ij}(\balpha,\bbeta)\,
\vx_i^\top\vy_j.
\label{eq:sclap_transport}
\end{equation}
Replacing CLAP's pooled similarity with
\[
s_{ab}=
s_{\mathrm{SCLAP\text{-}T}}
(\{\vx^{(a)}\}_{n_a},\{\vy^{(b)}\}_{m_b})/\tau
\]
in the symmetric InfoNCE loss in~(\ref{eq: infonce}) yields the SCLAP loss $\mathcal{L}_{\mathrm{SCLAP}}$. Thus,
SCLAP changes only the similarity function used during training, leaving the
dual-encoder architecture and batch-wise contrastive formulation unchanged. This change in similarity function nonetheless embeds a structural constraint.
The assumption of monotonicity is far more restrictive for this task than for ASR, where audio and transcript are genuinely order-aligned. When a clip contains several events, the caption may mention them in any order, or does not necessarily mention them all. This restriction however is mitigated by the fact that the plan  $\bgamma(\balpha,\bbeta)$ is fully determined by the
marginals, so the model's alignment freedom resides in $\balpha$ and
$\bbeta$ and assigning near-zero mass to positions that should not participate effectively removes them from the alignment. Our results suggest this is sufficient in practice.

Because $\balpha$ is computed from the audio sequence alone and $\bbeta$ from the text sequence alone, neither can use cross-modal information to decide which of its own positions matter. For the transport-weighted similarity to still separate matched from mismatched pairs during training, $\balpha$ and $\bbeta$ must therefore learn, from their own modality alone, to concentrate mass on the frames and tokens that carry the content described in a caption. Once trained, this means a weighted average under $\balpha$ and $\bbeta$ already concentrates each embedding on its relevant positions, without needing the transport plan itself. This yields a pooled similarity,
\begin{equation}
s_{\mathrm{SCLAP\text{-}P}}(\{\vx\}_n,\{\vy\}_m)
=
\cos\!\left(
\sum_i\alpha_i\vx_i,\,
\sum_j\beta_j\vy_j
\right),
\label{eq:sclap_pooled}
\end{equation}
which costs no more than a standard CLAP-style dot product. Crucially, since $\balpha$ and $\bbeta$ are each computed from a single modality, audio and text embeddings can be pooled and precomputed independently during inference, unlike $s_{\mathrm{SCLAP\text{-}T}}$, which requires both sequences at once to build $\bgamma$.
The same reasoning extends to localization. For temporal localization, we compare each individual audio frame against the pooled text embedding, yielding a per-frame score
\begin{equation}
\ell_i(\{\vx\}_n,\{\vy\}_m)
=
\vx_i^\top\sum_j\beta_j\vy_j,
\qquad i=1,\dots,n,
\label{eq:sclap_localization}
\end{equation}
which requires only the text marginal $\bbeta$. High values of $\ell_i$ indicate frames whose content aligns with the caption, giving a localization signal without any temporal supervision. We train with $s_{\mathrm{SCLAP\text{-}T}}$, and at inference compare $s_{\mathrm{SCLAP\text{-}T}}$ and $s_{\mathrm{SCLAP\text{-}P}}$ for retrieval, while using $\ell_i$ for localization.

\section{Experiments}
\subsection{Experimental Setup}
\label{sec:setup}
We evaluate SCLAP in short- and long-form settings. For short-form audio
(15--30\,s), we train on Clotho~\cite{drossos2019} and TACOS~\cite{primus2025tacos} datasets, with retrieval evaluated on Clotho and zero-shot localization evaluated on AudioSet-strong~\cite{hershey2021benefit}. 
For long-form audio (60--300\,s), we train and test retrieval and localization on the CASTELLA dataset~\cite{munakata2025castella}. In this case, training is done using only CASTELLA's global captions, while its local captions and timestamps are used only to evaluate zero-shot moment localization. \\

\noindent\textbf{Models.}
All models share the same dual-encoder backbone consisting of an ASiT spectrogram
transformer~\cite{Ahmed_2024} pretrained on AudioSet~\cite{audioset} and fine-tuned
with AudioSet-strong frame-level labels~\cite{schmid2024effectivepretrainingaudiotransformers}
(producing 1491 frame embeddings per clip) and
RoBERTa-base~\cite{DBLP:journals/corr/abs-1907-11692} for text, with
captions padded/masked to 77 tokens (maximum caption length 48). Both streams
are linearly projected to a shared 1024-dimensional space and
$\ell_2$-normalized per token. Models differ only in how the token sequences are converted into a similarity
score. SCLAP is trained using the transport readout
$s_{\mathrm{SCLAP\text{-}T}}$ defined in~\eqref{eq:sclap_transport}. At inference, we report both
$s_{\mathrm{SCLAP\text{-}T}}$ and the transport-free pooled readout
$s_{\mathrm{SCLAP\text{-}P}}$ defined in~\eqref{eq:sclap_pooled}, computed from the same checkpoint. We compare both SCLAP readouts with standard CLAP and TAP~\cite{xin2023improvingtextaudioretrievaltextaware}. TAP is an
attention-based CLAP variant in which the text \texttt{[CLS]} embedding
$t$ queries the audio frames. Given projected frame embeddings $x_i$, TAP
computes
{\small
\begin{equation}
\begin{gathered}
q=W_qt,\; k_i=W_kx_i,\; v_i=W_vx_i,\\[-2pt]
A_i=\operatorname{softmax}_i\!\left(
\frac{q^\top k_i}{\sqrt d}\right),\qquad
h=\operatorname{LN}\!\left(W_o\sum_i A_iv_i\right).
\end{gathered}
\end{equation}
}
The retrieval score is $s_{\mathrm{TAP}}=\cos(h,t)$ and is computed for each
candidate audio--text pair. For frame-level evaluation, we retain TAP's temporal structure using the
A$\times$V readout
\begin{equation}
(A\times V)_i
=
A_i\cos\!\left(\operatorname{LN}(W_o v_i),t\right).
\end{equation}
Attention alone indicates only relative frame selection, since its weights
sum to one even when no frame matches the query. A$\times$V additionally
measures whether each selected frame is semantically compatible with the
text. \\
\noindent\textbf{Training.}
All models are fine-tuned end-to-end for 20 epochs with AdamW and a cosine
schedule. The effective batch size is 16 on CASTELLA (batch size 4 with
4-step accumulation) and 32 on Clotho+TACOS (batch size 16 with 2-step
accumulation). For CLAP and SCLAP, we use 200 warm-up steps on CASTELLA and
120 on Clotho+TACOS. For TAP, 120 warm-up steps are used on all  datasets.
Hyperparameters are selected sequentially on the
validation split, with first selecting and fixing the contrastive temperature $\tau$, then the audio marginal
temperature $\theta_x$ for SCLAP, and finally the learning rate. The text marginal for SCLAP is set to $\theta_y=1$.
While $\tau$ is fixed for all approaches on all datasets, we use a trainable $\tau$ on CASTELLA for SCLAP to account for the longer recording durations. 
Code and models will be made publicly available upon acceptance. \\
%TAP is not included in this grid search and we use a fixed 512-dimensional attention head, $\tau=0.02$, a learning rate of $6\times10^{-5}$, with $\tau$ frozen during training.

\noindent\textbf{Evaluation.}
For audio--text \textbf{retrieval}, we report Recall@$k$, the percentage of
queries whose correct match appears among the top $k$ results, and mAP@10,
which also accounts for the ranking of relevant results within the top 10. \textbf{Localization} is evaluated zero-shot, without moment or timestamp
supervision during training. Because CASTELLA and AudioSet-strong define
localization differently, we use the standard metric for each task. CASTELLA
is formulated as ranked moment retrieval, i.e., we report Recall@$k$ at
intersection-over-union (IoU) thresholds $\{0.3,0.5,0.7\}$, where a prediction
is correct if one of the top-$k$ predicted moments overlaps the annotated
event by at least the given threshold. AudioSet-strong is instead a dense
multi-label sound-event detection task. We report global and macro-averaged
PSDS1, which summarize event-level detection across operating thresholds, and
pAUROC, which measures frame-level discrimination over a restricted
false-positive range. To measure \textbf{robustness to event dilution}, we additionally crop
CASTELLA clips with increasing amounts of context around the target event and
repeat the localization evaluation at each context length. All results are
reported as mean $\pm$ standard deviation over three training seeds.
\subsection{Short-form audio}

\textbf{Retrieval.} Table~\ref{tab:short-retrieval} shows short-form retrieval performance for CLAP, TAP, and the two SCLAP readouts, trained separately on Clotho and on Clotho+TACOS. All models are evaluated on the Clotho test set, a standard benchmark for retrieval in this field.
Whether models are trained on Clotho alone or on Clotho+TACOS, the performance differs little among the considered methods.
On Clotho, mAP@10 ranges from $29.25$ to
$29.79$, with TAP giving the highest point estimate. After training on
Clotho+TACOS, CLAP and the two SCLAP readouts obtain similar mAP@10
($30.96$, $30.88$, and $30.81$), while TAP reaches $30.10$. Most differences are comparable to the variation across seeds, and no method shows a consistent advantage across all retrieval metrics. \\

\noindent \textbf{Temporal localization.} 
Table~\ref{tab:clothotacos-sed} reports zero-shot localization performance of the models trained on Clotho+TACOS and evaluated on AudioSet-strong. It can be observed that SCLAP transfers better than CLAP, since the transport readout SCLAP-T improves pAUROC from 
$72.36$ to $76.13$ and PSDS1-macro from $12.57$ to $13.96$. The pooled readout SCLAP-P retains most of this improvement, showing that the
representations learned through alignment remain useful without computing
transport at inference. Using the A$\times$V frame-level readout previously defined, TAP retains some frame-ranking ability but transfers poorly to the localization task, reaching $61.59$ in pAUROC and near-zero PSDS1.

\begin{table}[t]
\centering
\footnotesize
\setlength{\tabcolsep}{3.5pt}
\caption{Retrieval performance on short-form audio, with models trained on Clotho or Clotho+TACOS and tested on Clotho test split (mean $\pm$ std over three seeds).}
\label{tab:short-retrieval}
\resizebox{\columnwidth}{!}{%
\begin{tabular}{lcccccccc}
\toprule
& \multicolumn{4}{c}{\textbf{Clotho}}
& \multicolumn{4}{c}{\textbf{Clotho+TACOS}} \\
\cmidrule(lr){2-5}\cmidrule(lr){6-9}
\textbf{Model}
& R@1 & R@5 & R@10 & mAP@10
& R@1 & R@5 & R@10 & mAP@10 \\
\midrule
CLAP
& \secondscore{17.66}{0.31}
& \score{44.95}{0.82}
& \score{59.36}{1.18}
& \score{29.25}{0.31}
& \secondscore{19.33}{0.57}
& \bestscore{46.70}{0.20}
& \secondscore{60.54}{0.24}
& \bestscore{30.96}{0.23} \\

TAP
& \bestscore{18.25}{0.67}
& \score{45.21}{0.62}
& \bestscore{59.94}{0.60}
& \bestscore{29.79}{0.68}
& \score{18.53}{0.15}
& \score{45.42}{0.27}
& \score{59.67}{0.24}
& \score{30.10}{0.11} \\

\midrule
SCLAP-P
& \score{17.49}{0.23}
& \bestscore{45.80}{0.18}
& \bestscore{59.94}{0.35}
& \secondscore{29.38}{0.27}
& \score{19.18}{0.59}
& \secondscore{46.69}{0.70}
& \bestscore{60.84}{0.38}
& \secondscore{30.88}{0.61} \\

SCLAP-T
& \score{17.53}{0.42}
& \secondscore{45.25}{0.55}
& \secondscore{59.66}{0.35}
& \score{29.31}{0.50}
& \bestscore{19.38}{0.66}
& \score{46.30}{0.28}
& \score{60.38}{0.32}
& \score{30.81}{0.54} \\
\bottomrule
\end{tabular}%
}
\end{table}
\begin{table}[t!]
\centering
\footnotesize
\caption{Zero-shot localization performance on short-form audio, with models trained on Clotho+Tacos and tested on AudioSet-strong (mean $\pm$ std over three seeds).}
\label{tab:clothotacos-sed}
\resizebox{0.8\columnwidth}{!}{
\begin{tabular}{lccc}
\toprule
\textbf{Model} & PSDS1-global & PSDS1-macro & pAUROC \\
\midrule
CLAP
    & \score{0.72}{0.18}
    & \score{12.57}{0.62}
    & \score{72.36}{1.18} \\
    \
TAP (A$\times$V)
    & \score{0.00}{0.00}
    & \score{0.46}{0.30}
    & \score{61.59}{4.26} \\
    \hline
SCLAP-P
    & \secondscore{1.27}{0.66}
    & \secondscore{13.24}{1.54}
    & \secondscore{75.61}{0.75} \\
SCLAP-T
    & \bestscore{1.43}{0.55}
    & \bestscore{13.96}{0.84}
    & \bestscore{76.13}{0.69} \\
\bottomrule
\end{tabular}}
\end{table}
\subsection{Long-form audio}
\textbf{Retrieval.}
Table~\ref{tab:castella-retrieval} presents long-form retrieval performance for CLAP, TAP, and the two SCLAP readouts trained and tested on CASTELLA. SCLAP-P yields the highest mean performance across all four metrics, reaching
$24.80$ mAP@10, compared with $23.86$ for TAP and $21.47$ for CLAP. The
difference between SCLAP and TAP remains within the variability across seeds, although both outperform uniform CLAP pooling on average. Beyond retrieval accuracy, SCLAP also has a practical advantage over TAP. Since its audio and text weights are computed independently, embeddings can be pooled and precomputed once, whereas TAP must recompute query-conditioned attention for every audio--text pair.
The transport and pooled SCLAP readouts also
perform similarly, indicating that the benefits of alignment training remain
useful with a simple pooled representation at inference.\\

\begin{table}[t]
\centering
\footnotesize
\caption{Retrieval performance on long-form audio, with models trained and tested on CASTELLA train/test splits (mean $\pm$ std
over three seeds).}
\label{tab:castella-retrieval}
\resizebox{0.7\columnwidth}{!}{
\begin{tabular}{lcccc}
\toprule
\textbf{Model} & R@1 & R@5 & R@10 & mAP@10 \\
\midrule
CLAP
& \score{11.64}{0.33}
& \score{34.07}{0.40}
& \score{49.21}{0.97}
& \score{21.47}{0.30} \\
TAP
& \secondscore{13.17}{0.99}
& \secondscore{37.99}{2.76}
& \score{52.33}{2.94}
& \score{23.86}{1.34} \\
\midrule
SCLAP-P
& \bestscore{13.49}{0.73}
& \bestscore{40.16}{1.30}
& \bestscore{54.66}{1.23}
& \bestscore{24.80}{0.91} \\
SCLAP-T
& \score{12.96}{0.56}
& \score{37.51}{2.69}
& \secondscore{54.07}{2.01}
& \secondscore{23.88}{0.98} \\
\bottomrule
\end{tabular}}
\end{table}
\begin{table}[t!]
\centering
\footnotesize
\setlength{\tabcolsep}{4pt}
\caption{Zero-shot localization performance on long-form audio, with models trained and tested on CASTELLA train/test splits (mean $\pm$ std over three seeds).
%with up to 320\,s of context,
CLAP and SCLAP use frame--text cosine similarity, whereas TAP uses attention-weighted value--text compatibility. No transport is solved at inference for SCLAP.}
\label{tab:castella-localization}
\resizebox{0.9\columnwidth}{!}{
\begin{tabular}{lcccccc}
\toprule
& \multicolumn{3}{c}{\textbf{R@1}}
& \multicolumn{3}{c}{\textbf{R@5}} \\
\cmidrule(lr){2-4}\cmidrule(lr){5-7}
\textbf{Model}
& IoU$\geq$.3 & IoU$\geq$.5 & IoU$\geq$.7
& IoU$\geq$.3 & IoU$\geq$.5 & IoU$\geq$.7 \\
\midrule
CLAP
& \secondscore{6.10}{0.47}
& \score{2.03}{0.10}
& \score{0.60}{0.10}
& \secondscore{15.50}{0.63}
& \secondscore{8.86}{0.21}
& \secondscore{3.59}{0.31} \\
TAP (A$\times$V)
& \score{4.67}{0.78}
& \secondscore{2.21}{0.58}
& \secondscore{0.78}{0.37}
& \score{13.64}{1.71}
& \score{7.66}{1.40}
& \score{2.75}{0.73} \\
\hline
SCLAP
& \bestscore{7.60}{1.05}
& \bestscore{2.87}{0.78}
& \bestscore{0.96}{0.37}
& \bestscore{17.00}{0.85}
& \bestscore{8.98}{0.64}
& \bestscore{3.71}{0.10} \\
\bottomrule
\end{tabular}}
\end{table}
\begin{table}[t!]
\centering
\footnotesize
\setlength{\tabcolsep}{5pt}
\caption{Moment retrieval R@1 (IoU$\geq$0.3) for models trained on CASTELLA as each test clip is cropped to an
increasingly wider context window around the target event (mean $\pm$ std over three seeds). }
\label{tab:context-sweep}
\resizebox{0.8\columnwidth}{!}{\begin{tabular}{lccccc}
\toprule
\textbf{Context (s)} & 20 & 40 & 80 & 160 & 320 \\
\midrule
CLAP
& \score{15.62}{1.08}
& \secondscore{12.27}{0.37}
& \secondscore{8.26}{0.31}
& \secondscore{6.88}{0.37}
& \secondscore{6.10}{0.47} \\
TAP (A$\times$V)
& \bestscore{18.25}{0.10}
& \score{11.37}{0.73}
& \score{7.12}{0.37}
& \score{5.57}{0.82}
& \score{4.67}{0.78} \\
\hline
SCLAP
& \secondscore{17.12}{1.35}
& \bestscore{12.69}{0.83}
& \bestscore{10.53}{0.10}
& \bestscore{9.40}{0.72}
& \bestscore{7.60}{1.05} \\
\bottomrule
\end{tabular}}
\end{table}
\noindent\textbf{Temporal localization.}
Long-form temporal localization is the setting in which SCLAP shows its
clearest advantage. Table~\ref{tab:castella-localization} reports zero-shot moment retrieval performance on full CASTELLA recordings, without
moment-level supervision during training.
Table~\ref{tab:context-sweep} reports the performance as each test clip is cropped to an increasingly wide context window around the target event.
SCLAP obtains the highest mean performance on all six full-context
metrics in Table~\ref{tab:castella-localization}.
In Table~\ref{tab:context-sweep}, TAP performs best at $20$\,s, where little unrelated audio surrounds the event, whereas SCLAP performs best at every context length from $40$ to $320$\,s. At $320$\,s, SCLAP reaches $7.60$ R@1 at IoU$\geq0.3$, compared with $6.10$ for CLAP and $4.67$ for TAP. TAP is therefore effective in short contexts, while the ordered alignment learned by SCLAP is more robust when the target event is diluted by long stretches of unrelated audio.

\section{Conclusion}

We introduced SCLAP, which replaces global mean pooling during training with
a one-dimensional OT  alignment between audio frames and text
tokens. On short-form recordings, SCLAP remains competitive with CLAP and TAP for
retrieval. On long-form recordings, its pooled readout gives the
highest mean retrieval performance, while its clearest advantage appears in
zero-shot temporal localization. SCLAP becomes increasingly robust relative
to the considered baselines as more unrelated context surrounds the target event.
Moreover, the transport plan does not need to be computed at inference, with pooled and
frame-level readouts retaining the benefits of sequence-level alignment learned during training. Future work will evaluate the benefit of this framework  for localizing pathological speech segments in clinical speech processing.

\bibliographystyle{IEEEtran}
\bibliography{refs}

@inproceedings{zhao2025superclip,
title={Super{CLIP}: {CLIP} with Simple Classification Supervision},
author={Weiheng Zhao and Zilong Huang and Jiashi Feng and Xinggang Wang},
booktitle={Proc. Annual Conference on Neural Information Processing Systems},
year={2025},
month = {Dec.},
address = {San Diego, USA},
}

@INPROCEEDINGS{primus2025tacos,
  author={Primus, Paul and Schmid, Florian and Widmer, Gerhard},
  booktitle={Proc. IEEE Workshop on Applications of Signal Processing to Audio and Acoustics}, 
  title={{TACOS: T}emporally-aligned Audio CaptiOnS for Language-Audio Pretraining}, 
  year={2025},
  address = {California, USA},
  month = {Oct.},
}

@inproceedings{radford2021,
  author    = {Radford, Alec and Kim, Jong Wook and Hallacy, Chris and Ramesh, Aditya and Goh, Gabriel and Agarwal, Sandhini and Sastry, Girish and Askell, Amanda and Mishkin, Pamela and Clark, Jack and Krueger, Gretchen and Sutskever, Ilya},
  title     = {Learning Transferable Visual Models From Natural Language Supervision},
  booktitle = {Proc. International Conference on Machine Learning},
  pages     = {8748--8763},
  year      = {2021},
  address = {Virtual},
  month = {July},
}

@inproceedings{jia2021,
  author    = {Chao Jia and Yinfei Yang and Ye Xia and Yi-Ting Chen and Zarana Parekh and Hieu Pham and Quoc V. Le and Yun-Hsuan Sung and Zhen Li and Tom Duerig},
  title     = {Learning Transferable Visual Models From Natural Language Supervision},
  booktitle = {Proc. International Conference on Machine Learning},
  pages     = {4904--4916},
  year      = {2021},
  address = {Virtual},
  month = {July},
}

@INPROCEEDINGS{guzhov2021audioclipextendingclipimage,
  author={Guzhov, Andrey and Raue, Federico and Hees, Jörn and Dengel, Andreas},
  booktitle={Proc. IEEE International Conference on Acoustics, Speech and Signal Processing}, 
  title={{AudioCLIP:} {E}xtending {CLIP} to Image, Text and Audio}, 
  year={2022},
  pages={976--980},
  month = {May},
  address = {Singapore},
}

@INPROCEEDINGS{elizalde2022claplearningaudioconcepts,
  author={Elizalde, Benjamin and Deshmukh, Soham and Ismail, Mahmoud Al and Wang, Huaming},
  booktitle={Proc. IEEE International Conference on Acoustics, Speech and Signal Processing}, 
  title={{CLAP: L}earning Audio Concepts from Natural Language Supervision}, 
  year={2023},
  address = {Rhodes Island, Greece},
  month = {June},
}

@INPROCEEDINGS{wu2024largescalecontrastivelanguageaudiopretraining,
  author={Wu, Yusong and Chen, Ke and Zhang, Tianyu and Hui, Yuchen and Berg-Kirkpatrick, Taylor and Dubnov, Shlomo},
  booktitle={Proc. IEEE International Conference on Acoustics, Speech and Signal Processing}, 
  title={Large-Scale Contrastive Language-Audio Pretraining with Feature Fusion and Keyword-to-Caption Augmentation}, 
  year={2023},
  month = {June},
  address = {Rhodes Island, Greece},
}

@INPROCEEDINGS{yuan2024tc,
  author={Yuan, Yi and Chen, Zhuo and Liu, Xubo and Liu, Haohe and Xu, Xuenan and Jia, Dongya and Chen, Yuanzhe and Plumbley, Mark D. and Wang, Wenwu},
  booktitle={Proc. IEEE International Workshop on Machine Learning for Signal Processing}, 
  title={{T-CLAP: T}emporal-Enhanced Contrastive Language-Audio Pretraining}, 
  year={2024},
  month = {Sept.},
  address = {London, UK}
  
}

@misc{kaloga2025differentiable,
  title         = {A Differentiable Alignment Framework for Sequence-to-Sequence
                   Modeling via Optimal Transport},
  author        = {Kaloga, Yacouba and Kumar, Shashi and Motlicek, Petr and Kodrasi, Ina},
  year          = {2025},
howpublished = {arXiv preprint},
url           = {https://arxiv.org/abs/2502.01588},
}

@INPROCEEDINGS{munakata2025lan,
  author={Munakata, Hokuto and Nishimura, Taichi and Nakada, Shota and Komatsu, Tatsuya},
  booktitle={Proc. IEEE International Conference on Acoustics, Speech and Signal Processing}, 
  title={Language-based Audio Moment Retrieval}, 
  year={2025},
  month = {Apr.},
  address = {Hyderabad, India},
}

@InProceedings{oncescu2024dis,
  author    = {Oncescu, Andreea-Maria and Henriques, Jo{\~a}o F. and Koepke, A. Sophia},
  title     = {Dissecting Temporal Understanding in Text-to-Audio Retrieval},
  booktitle = {Proc. ACM Multimedia Conference },
  year      = {2024},
  month = {Nov.},
  address = {Ottawa, Canada},
}

@inproceedings{li2024mga,
  title={Advancing Multi-Grained Alignment for Contrastive Language-Audio Pre-Training},
  author={Li, Yiming and Guo, Zhifang and Wang, Xiangdong and Liu, Hong},
  booktitle={Proc. ACM International Conference on Multimedia},
  pages={7356--7365},
  year={2024},
  month = {Nov.},
  address = {Melbourne, Australia},
}

@inproceedings{li2026fine,
    title = "{F}ine{LAP}: {T}aming Heterogeneous Supervision for Fine-grained Language-Audio Pretraining",
    author = "Li, Xiquan and Xu, Xuenan and Ma, Ziyang and Chen, Wenxi and He, Haolin and Kong, Qiuqiang and Chen, Xie",
    booktitle = "Proc. Annual Meeting of the Association for Computational Linguistics",
    month = {Jan.},
    year = "2026",
    address = "San Diego, USA",
    pages = "10393--10408"
}

@INPROCEEDINGS{wu2024collap,
  author={Wu, Junda and Li, Warren and Novack, Zachary and Namburi, Amit and Chen, Carol and McAuley, Julian},
  booktitle={Proc. IEEE International Conference on Acoustics, Speech and Signal Processing}, 
  title={{CoLLAP: C}ontrastive Long-form Language-Audio Pretraining with Musical Temporal Structure Augmentation}, 
  year={2025},
  month = {Apr.},
  address = {Hyderabad, India}
}

@inproceedings{
luong2024revisiting,
title={Revisiting Deep Audio-Text Retrieval Through the Lens of Transportation},
author={Manh Luong and Khai Nguyen and Nhat Ho and Reza Haf and Dinh Phung and Lizhen Qu},
booktitle={Proc. International Conference on Learning Representations},
year={2024},
address = {Vienna, Austria},
month = {May},
}

@InProceedings{pmlr-v235-shi24b,
  title = 	 {{OT}-{CLIP}: Understanding and Generalizing {CLIP} via Optimal Transport},
  author =       {Shi, Liangliang and Fan, Jack and Yan, Junchi},
  booktitle = 	 {Proc. International Conference on Machine Learning},
  pages = 	 {44865--44886},
  year = 	 {2024},
  month = 	 {July},
  address = {Vienna, Austria}
}

@INPROCEEDINGS{drossos2019,
  author={Drossos, Konstantinos and Lipping, Samuel and Virtanen, Tuomas},
  booktitle={Proc. IEEE International Conference on Acoustics, Speech and Signal Processing}, 
  title={Clotho: an Audio Captioning Dataset}, 
  year={2020},
  pages={736--740},
  month = {May},
  address = {Barcelona, Spain}
}

@misc{DBLP:journals/corr/abs-1907-11692,
  title        = {{RoBERTa}: A Robustly Optimized {BERT} Pretraining Approach},
  author       = {Liu, Yinhan and Ott, Myle and Goyal, Naman and Du, Jingfei and
                  Joshi, Mandar and Chen, Danqi and Levy, Omer and Lewis, Mike and
                  Zettlemoyer, Luke and Stoyanov, Veselin},
  year         = {2019},
  howpublished = {arXiv preprint},
  url          = {https://arxiv.org/abs/1907.11692},
}

@ARTICLE{Ahmed_2024,
  author={Ahmed, Sara Atito Ali and Awais, Muhammad and Wang, Wenwu and Plumbley, Mark D. and Kittler, Josef},
  journal={IEEE/ACM Transactions on Audio, Speech, and Language Processing}, 
  title={{ASiT: L}ocal-Global Audio Spectrogram Vision Transformer for Event Classification}, 
  year={2024},
  volume={32},
  pages={3684--3693},
}

@INPROCEEDINGS{schmid2024effectivepretrainingaudiotransformers,
  author={Schmid, Florian and Morocutti, Tobias and Foscarin, Francesco and Schlüter, Jan and Primus, Paul and Widmer, Gerhard},
  booktitle={Proc. IEEE International Conference on Acoustics, Speech and Signal Processing}, 
  title={Effective Pre-Training of Audio Transformers for Sound Event Detection}, 
  year={2025},
  month = {Apr.},
  address = {Hyderabad, India}
}

@INPROCEEDINGS{xin2023improvingtextaudioretrievaltextaware,
  author={Xin, Yifei and Yang, Dongchao and Zou, Yuexian},
  booktitle={Proc. IEEE International Conference on Acoustics, Speech and Signal Processing}, 
  title={Improving Text-Audio Retrieval by Text-Aware Attention Pooling and Prior Matrix Revised Loss}, 
  year={2023},
  month = {June},
  address = {Rhodes Island, Greece},
}

@INPROCEEDINGS{munakata2025castella,
  author={Munakata, Hokuto and Imamura, Takehiro and Nishimura, Taichi and Komatsu, Tatsuya},
  booktitle={Proc. IEEE International Conference on Acoustics, Speech and Signal Processing}, 
  title={{CASTELLA: L}ong Audio Dataset with Captions and Temporal Boundaries}, 
  year={2026},
  pages={15352--15356},
  address = {Barcelona, Spain},
  month = {May},
}

@article{peyré2020computationaloptimaltransport,
  title   = {Computational Optimal Transport: With Applications to Data Science},
  author  = {Peyr{\'e}, Gabriel and Cuturi, Marco},
  journal = {Foundations and Trends in Machine Learning},
  volume  = {11},
  number  = {5--6},
  pages   = {355--607},
  year    = {2019},
}

@inproceedings{hershey2021benefit,
  author    = {Hershey, Shawn and Ellis, Daniel P. W. and Fonseca, Eduardo and Jansen, Aren and Liu, Caroline and Moore, R. Channing and Plakal, Manoj},
  title     = {The Benefit of Temporally-Strong Labels in Audio Event Classification},
  booktitle = {Proc. IEEE International Conference on Acoustics, Speech and Signal Processing},
  year      = {2021},
  pages     = {366--370},
  month = {June},
  address = {Toronto, Canada}
}

@inproceedings{audioset,
  author    = {Gemmeke, Jort F. and Ellis, Daniel P. W. and Freedman, Dylan and Jansen, Aren and Lawrence, Wade and Moore, R. Channing and Plakal, Manoj and Ritter, Marvin},
  title     = {Audio Set: An Ontology and Human-Labeled Dataset for Audio Events},
  booktitle = {Proc. IEEE International Conference on Acoustics, Speech and Signal Processing},
  year      = {2017},
  pages     = {776--780},
  month = {Mar.},
  address = {New Orleans, USA},
  doi       = {10.1109/ICASSP.2017.7952261}
}

\end{document}